\documentclass[11pt,a4paper]{article} 
\usepackage{style}

\usepackage{amsmath,amssymb,epsfig,bm,amsfonts,yfonts,bbm,mathrsfs,hyperref}
\usepackage{dsfont,mathtools,graphicx,empheq,color,tikz,stmaryrd,gensymb,braket}
\usepackage{dynkin-diagrams}

\definecolor{rot}{rgb}{0.7,0,0}
\definecolor{gruen}{rgb}{0,0.7,0}
\definecolor{blau}{rgb}{0,0,0.6}

\title{Information geometry of Euclidean quantum fields}

\author{Stefan Floerchinger}

\affiliation{Theoretisch-Physikalisches Institut, Max-Wien Platz 1, 07743 Jena, Germany}

\emailAdd{stefan.floerchinger@uni-jena.de}

\abstract{Information geometry provides differential geometric concepts like a Riemannian metric, connections and covariant derivatives on spaces of probability distributions. We discuss here how these concepts apply to quantum field theories in the Euclidean domain which can also be seen as statistical field theories. The geometry has a dual affine structure corresponding to sources and field expectation values seen as coordinates. A key concept is a new generating functional, which is a functional generalization of the Kullback-Leibler divergence. From its functional derivatives one can obtain connected as well as one-particle irreducible correlation functions. It also encodes directly the geometric structure, i.\ e.\ the Fisher information metric and the two dual connections, and it determines asymptotic probabilities for field configurations through Sanov's theorem.}

\begin{document}

\maketitle

\section{Introduction}

Information theoretic aspects of quantum field theory are getting more important recently, partly triggered by the advancement of quantum information theory. Traditionally this has been discussed in the context of black holes and the information paradox \cite{Bombelli:1986rw,Srednicki:1993im,Callan:1994py,Witten:2018zxz}, but also in the context of non-equilibrium dynamics and thermalization \cite{Floerchinger:2020ogh, Dowling:2020nxc}, or for expanding quantum fields \cite{bergesDynamicsEntanglementExpanding2018,bergesThermalExcitationSpectrum2018, giantsosEntanglementEntropyFourdimensional2022,boutivasEntanglementExpansion2023}. Information theoretic methods can be used to classify states, for example in terms of entropy or entanglement entropy, to compare states in terms of relative entropies etc. 

It is interesting in that context that there is actually a sophisticated geometric structure on the space of probability distributions or quantum density matrices. This has been developed mathematically in the field of information geometry, see refs.\ \cite{amariInformationGeometryIts2016,ayInformationGeometry2017} for recent reviews. More specific, the Fisher information metric, or its quantum generalization, the quantum Fisher information metric, provides a natural Riemannian metric. Beyond this, there is also a natural connection available, or actually two that are dual, described by the Amari-Chentsov structure \cite{amariInformationGeometryIts2016,ayInformationGeometry2017}. They differ from the Levi-Civita connection and are not metric-compatible or, in other words, they have non-metricity. Even more, the entire geometric structure is very nicely encoded in terms of divergences such as the Kullback-Leibler divergence.

Using these geometric structures to better understand quantum field dynamics is an excellent perspective for the coming years. As a preparation for this we follow a related program in the present article: we study Euclidean quantum field theories that directly have a probability interpretation, or that can be seen as classical statistical field theories with one more spatial dimension, from an information theoretic point of view. This has the advantage that concepts can be taken over from the mathematical literature on classical information geometry directly, with the only generalization being from functions to functionals\footnote{Even that step has been partly taken by mathematicians already \cite{ayInformationGeometry2017}.}. 

In other words, in the present work we discuss the information geometry of statistical field theories which can also be seen as quantum field theories in the Euclidean domain. We explore specifically how source fields or field expectation values can be seen as alternative coordinates in the space of functional probability distributions and what kind of geometric structures are associated to them. The Riemannian Fisher information metric will be seen to correspond to connected two-point correlation functions. The two dual connections are corresponding to different kinds of three-point correlation functions (connected and one-particle irreducible).

Let us mention here that somewhat different information theoretic concepts (based on the Wasserstein metric) have been applied in particle physics previously \cite{komiskeMetricSpaceCollider2019,komiskeHiddenGeometryParticle2020}. Moreover, in ref.~\cite{erdmengerInformationGeometryQuantum2020}, Erdmenger, Grosvenor and Jefferson explored the use of the quantum Fisher information metric in different model systems. In ref.\ \cite{erdmengerQuantifyingInformationFlows2022} they studied relative entropy and the connections of neural networks with the renormalization group, see also ref.\ \cite{grosvenorEdgeChaosQuantum2022}.

An information-theoretic view on the renormalization group is actually being developed since many years, see refs.\ \cite{diosiMetricizationThermodynamicstateSpace1984, gaiteFieldTheoryEntropy1996,
dolanRenormalisationGroupFlow1998,
brodySymmetryRealspaceRenormalisation1998, dolanRenormalizationGroupFlow1999,
preskillQuantumInformationPhysics2000, casiniTheoremEntanglementEntropy2007, apenkoInformationTheoryRenormalization2012, benyInformationgeometricApproachRenormalization2015, benyRenormalizationGroupStatistical2015, fowlerMisanthropicEntropyRenormalization2022,koenigsteinNumericalFluidDynamics2022}.
Recently also connections between the renormalization group and optimal transport have been explored \cite{cotlerRenormalizationGroupFlow2023, bermanInverseExactRenormalization2022}.

Relative entropy and the Fisher information metric have been discussed as an interesting possibility to distinguish between different quantum field theories \cite{balasubramanianRelativeEntropyProximity2015}. In this context, infinite information theoretic distances have been discussed in refs.\ \cite{stoutInfiniteDistanceLimits2021, basileEmergentStringsInfinite2022, stoutInfiniteDistancesFactorization2022} with view on applications in the context of the Swampland program, see refs.\ \cite{brennanStringLandscapeSwampland2018,paltiSwamplandIntroductionReview2019} for reviews.

The quantum Fisher information has also been employed in the context of entanglement detection, see refs.\ \cite{strobelFisherInformationEntanglement2014,haukeMeasuringMultipartiteEntanglement2016,pezzeQuantumMetrologyNonclassical2018} and references therein. It was also discussed in the context of holography, see e.\ g.\ ref.\ \cite{banerjeeConnectingFisherInformation2018}. In refs. \cite{aokiGeometriesFieldTheories2015, aokiFlowEquationLarge2016} an emergent metric in large-$N$ scalar theories has been discussed. Similar concepts have also been used to quantify complexity \cite{chapmanDefinitionComplexityQuantum2018,erdmengerExactGravityDuals2022,erdmengerComplexityGeometryHolographic2022}.

Geometric concepts also find applications in thermodynamics, see \cite{ruppeinerRiemannianGeometryThermodynamic1995, brodyInformationGeometryVapour2009} for reviews or to characterize phase transitions \cite{carolloGeometryQuantumPhase2020}. Quantum states have a geometric characterization as well, see for example ref.\ \cite{bengtssonGeometryQuantumStates2017} for an introduction. Relative entropy has recently been used to formulate entropic uncertainty relations for quantum field theories \cite{floerchingerRelativeEntropicUncertainty2022}.

Let us remark that discussion we present here is from a mathematical point of view heuristic. For example, issues with different topologies or the definition of the functional integral will not be discussed. We leave in particular the issue of renormalization to future work.

This paper is organized as follows. In section \ref{sec:EuclideanQuantumFieldTheory} we briefly recall the conceptual setup of Euclidean quantum field theory before discussing a natural affine structure in terms of sources in \ref{sec:AffineGeometryForSources}. In \ref{sec:TheFisherInformationMetric} we introduce the Fisher information metric and discuss its significance in the field theoretic context. In section \ref{sec:ExpectationValuesAsCoordinates} we introduce expectation value fields as alternative coordinates using Legendre transforms and the quantum effective action. The corresponding affine geometry is discussed in section \ref{sec:AffineGeometryForExpectationValues}. In section \ref{sec:DivergenceFunctional} we discuss a new generating functional for quantum field theory, the functional generalization of the Kullback-Leibler divergence in terms of source coordinates, expectation value coordinates and combinations thereof. Also the information theoretic significance of this object will be discussed. Subsequently, in section \ref{sec:FunctionalIntegralRepresentationForDivergenceFunctional} we present functional integral representations of the divergence functional including steepest descend or one-loop approximations to it. In section \ref{sec:Connections} we discuss different connections in more detail, before explaining how they can be obtained from the divergence functional in general coordinates in section \ref{sec:MetricAndConnectionFormDivergence}. Finally, we draw some conclusions in section \ref{sec:Conclusions}.

%%%%%%%%%%%%
%%%%%%%%%%%%

%%%%%%%%%%%%
%%%%%%%%%%%%

\section{Euclidean quantum field theory}
\label{sec:EuclideanQuantumFieldTheory}

We study here a bosonic quantum field theory in Euclidean spacetime, or statistical field theory, with the partition function
\begin{equation}
  Z[J] = e^{W[J]} = \int D\phi \exp\left( -S[\phi] + \int_x \sum_n J_n(x) \phi_n(x) \right).
\end{equation}
The index $n$ labels field components, where the fields could be fundamental or composite, and we use the abbreviation
\begin{equation}
  \int_x = \int d^d x \sqrt{g(x)}
\end{equation}
with $g(x) = \text{det}(g_{\mu\nu}(x))$ the determinant of the Euclidean spacetime metric.

From the partition function $Z[J]$, or the Schwinger functional $W[J]$ (known as cumulant generating function in statistics) one easily derives expectation values, correlation functions or other interesting observables. One can easily show, using Hölder's inequality, that $W[J]$ is convex,
\begin{equation}
  W[(1-t)J^\prime + tJ^{\prime\prime}] \leq (1-t) W[J^\prime] + t W[J^{\prime\prime}],
\end{equation}
with $0\leq t\leq 1$. 

In the following, it will also be convenient to use an abstract index which combines position and field component, $\alpha = (x, n)$, and to use Einsteins summation convention,
\begin{equation}
  J^\alpha \phi_\alpha = \int_x \sum_n J_n(x) \phi_n(x) ,
\end{equation}
with $J^\alpha = J_n(x)$ and $\phi_\alpha = \phi_n(x)$. It is usually clear from the context how expressions involving abstract indices can be made concrete.

A Euclidean quantum field theory, or a statistical field theory, for example to describe critical phenomena, is a probabilistic theory where the random variables are the field configurations $\phi_n(x)$, and a probability density with respect to the functional integral measure $D\phi$ is given by
\begin{equation}
  p[\phi, J] = \exp\left( -S[\phi] + J^\alpha \phi_\alpha - W[J] \right).
  \label{eq:defFunctProbDistribution}
\end{equation}
The source field $J^\alpha$ can be seen as a parameter within a class of probability distributions.

One may generalize this setup somewhat, in the sense that the fundamental random variables could be some other microscopic degrees of freedom (e.\ g.\ of a lattice model or similar), which we call $\chi$ but do not specify them further. The probability density with respect to the measure $D\chi$ is then
\begin{equation}
  p[\chi, J] = \exp\left( -I[\chi] + J^\alpha \phi_\alpha[\chi] - W[J] \right),
  \label{eq:ProbabilityDistributionExponentialFamily}
\end{equation}
with
\begin{equation}
  Z[J] = e^{W[J]} = \int D\chi \exp\left( -I[\chi] + J^\alpha \phi_\alpha[\chi] \right).
  \label{eq:SchwingerFunctionalChi}
\end{equation}
The fields $\phi_\alpha[\chi]$ are now some functionals of the fundamental or microscopic random variables $\chi$. The functional $I[\chi]$ plays the role of the action $S[\phi]$ but can differ as a result of the variable change with fixed measure
\begin{equation}
  D\chi \exp\left( - I[\chi] \right) = D\phi \exp\left( -S[\phi] \right).
  \label{eq:chiphiCoordinateChange}
\end{equation}
A class of probability distributions as in eq.\ \eqref{eq:ProbabilityDistributionExponentialFamily} with the sources $J^\alpha$ seen as parameters, is known as an exponential family in information geometry \cite{amariInformationGeometryIts2016,ayInformationGeometry2017}.

Note that eq.\ \eqref{eq:ProbabilityDistributionExponentialFamily} together with \eqref{eq:SchwingerFunctionalChi} permits also to take generalized coupling constants to be part of the set $J^\alpha$. Indeed, the operators $\phi_\alpha[\chi]$ are not restricted to be fundamental fields and could also be composite.

\section{Affine geometry for sources}
\label{sec:AffineGeometryForSources}
Let us note here immediately that there is an affine structure on this exponential family in the sense that an affine transformation of sources,
\begin{equation}
  J^\alpha \to J^{\prime\alpha} = M^\alpha_{~\beta} J^\beta + c^\alpha,
  \label{eq:AffineTransformationsSources}
\end{equation}
with invertible $M^\alpha_{~\beta}$, leads again to a probability distribution of the form \eqref{eq:ProbabilityDistributionExponentialFamily}, i.\ e.\ in the exponential family. Note that this is not the case for non-linear transformation of sources.

In a related way, affine transformations of the form \eqref{eq:AffineTransformationsSources} preserve the convexity of the Schwinger functional $W[J]$, while more general non-linear transformation do not.

Moreover, the affine structure allows connecting probability distributions associated to the sources $J^{\prime\alpha}$ and $J^{\prime\prime\alpha}$ through a distribution with source $J^\alpha(t)$ given by the flat, so-called $e$-geodesic\footnote{A geodesic is here a line connecting two points, determined by a connection, and not necessarily the shortest path in any sense.},
\begin{eqnarray}
  J^\alpha(t) = (1-t) J^{\prime\alpha} + t J^{\prime\prime \alpha},
  \label{eq:DefEGeodesicEndpoints}
\end{eqnarray}
where $0\leq t \leq 1$. Geometrically this means that the manifold of probability distributions in the exponential family has a flat connection in terms of the coordinates $J^\alpha$. As trajectories, $e$-geodesics are characterized by the differential equation
\begin{equation}
  \frac{d^2}{dt^2} J^\alpha(t) + (\Gamma_\text{E})_{\beta~\gamma}^{\phantom{\beta}\alpha}[J] \, \frac{d}{dt} J^\beta(t) \, \frac{d}{dt} J^\gamma(t) = 0,
\end{equation}
where the connection vanishes here, in terms of source coordinates, 
\begin{equation}
  (\Gamma_\text{E})_{\beta~\gamma}^{\phantom{\beta}\alpha}[J] = 0.  
  \label{eq:EConnectionSourceCoordinates}
\end{equation}
In another coordinate system, related to $J^\alpha$ in a non-linear way, this would not be the case, however.

\section{The Fisher information metric}
\label{sec:TheFisherInformationMetric}

Spaces of probability distributions have a natural Riemannian metric, the Fisher information metric. For distributions parametrized by coordinates $J^\alpha$ it is given by
\begin{equation}
  \begin{split}
    G_{\alpha\beta}[J] = & \int D\chi \, p[\chi, J]\, \frac{\delta}{\delta J^\alpha} \ln p[\chi,J] \, \frac{\delta}{\delta J^\beta} \ln p[\chi,J] \\
    = & - \int D\chi \, p[\chi, J] \, \frac{\delta^2}{\delta J^\alpha \delta J^\beta} \ln p[\chi, J].
  \end{split}
\end{equation}
In the last equation we have used the product rule and that $\int D\chi \, p[\chi, J] = 1$.

The information-theoretic significance of the Fisher metric is that the infinitesimal Fisher-Rao distance of two nearby probability distributions
\begin{equation}
  ds^2 = G_{\alpha\beta}[J] d J^\alpha dJ^\beta
\end{equation}
gives a measure for how well distributions at $J$ and $J^\prime$ can be distinguished \cite{coverElementsInformationTheory2006}. As usual, in Riemannian geometry, the length of a path is defined through line integrals of $ds$.

For the exponential family, one finds easily
\begin{equation}
  G_{\alpha\beta}[J] = \frac{\delta^2}{\delta J^\alpha \delta J^\beta} W[J] = \langle \phi_\alpha[\chi] \phi_\beta[\chi] \rangle - \langle \phi_\alpha[\chi] \rangle \langle \phi_\beta[\chi] \rangle.
  \label{eq:FisherMetricSourceCoordinates}
\end{equation}
The Fisher metric agrees with the connected correlation function of fields! With the exception of Gaussian theories, this correlation function depends still on the source $J^\alpha$ and the metric is therefore not constant. Let us note that the metric \eqref{eq:FisherMetricSourceCoordinates} can be seen as an extension of Zamolodchikov's metric defined for conformal field theories \cite{Zamolodchikov:1986gt}, see \cite{balasubramanianRelativeEntropyProximity2015, stoutInfiniteDistanceLimits2021} for further discussion.

We are now in the interesting situation that the space of probability distributions as parametrized by $J^\alpha$ has a non-trivial Riemannian metric $G_{\alpha\beta}[J]$, but at the same time has an affine structure with $e$-geodesics corresponding to vanishing connection, $(\Gamma_\text{E})_{\beta~\gamma}^{\phantom{\beta}\alpha}[J] = 0$. It is clear that this connection is, for non-Gaussian theories, not the Levi-Civita connection corresponding to the metric $G_{\alpha\beta}[J]$, and it remains to characterize it further.

\section{Expectation values as coordinates}
\label{sec:ExpectationValuesAsCoordinates}

Instead of the sources $J^\alpha$ one can parametrize the probability distributions for $\chi$ or the fields $\phi_\alpha[\chi]$ in terms of the expectation values
\begin{equation}
  \Phi_\alpha = \langle \phi_\alpha[\chi] \rangle = \frac{\delta}{\delta J^\alpha} W[J] = \int D\chi \, p[\chi, J] \, \phi_\alpha[\chi].
  \label{eq:DefExpectaionValues}
\end{equation}
With the exception of Gaussian theories this is a non-linear change of coordinates from the sources $J^\alpha$.

One way to describe this change of variables is in terms of Legendre transforms. The quantum effective action or one-particle irreducible effective action is defined as
\begin{equation}
  \Gamma[\Phi] = \sup_J \left( J^\alpha \Phi_\alpha - W[J] \right),  
  \label{eq:DefEffectiveAction}
\end{equation}
and as a Legendre transform it is a convex functional of expectation values $\Phi_\alpha$. In the context of the theory of large deviations this is known as the Cram\'{e}r function \cite{demboLargeDeviationsTechniques2010}. One can also see $\Gamma[\Phi]$ as the negative of the infimum of the differential information entropy,
\begin{equation}
  \Gamma[\Phi] = - \inf_{J} \left( - \int D\chi \, p[\chi,J] \, \ln p[\chi, J] \right),
\end{equation}
for given expectation value $\Phi_\alpha$. Note that as a differential entropy this is not necessarily positive. The quantum effective action satisfies the field equation
\begin{equation}
  \frac{\delta}{\delta \Phi_\alpha} \Gamma[\Phi] = J^\alpha.
  \label{eq:FieldEquationEffectiveAction}
\end{equation}
With these relations one can write the probability density for $\chi$ as
\begin{equation}
  p[\chi, \Phi] = \exp\left( - I[\chi] + \frac{\delta \Gamma[\Phi]}{\delta \Phi_\alpha} (\phi_\alpha - \Phi_\alpha) + \Gamma[\Phi] \right).
  \label{eq:FunctionalProbabilityExpectationValues}
\end{equation}
The normalization condition for the distribution \eqref{eq:FunctionalProbabilityExpectationValues}, $\int D\chi \, p[\chi, \Phi]=1$, gives the well known background identity for $\Gamma[\Phi]$. On the other side, eq.\ \eqref{eq:DefExpectaionValues} is a linear relation between $\Phi_\alpha$ and $p$, and it should therefore be possible to write $p[\chi, \Phi]$ as a linear functional in $\Phi_\alpha$. 

To investigate this in more detail we first define $\Phi^\text{eq}_\alpha$ as the expectation value at vanishing source, i.\ e.\ the solution of eq.\ \eqref{eq:FieldEquationEffectiveAction} at $J^\alpha=0$. One can infer immediately that
\begin{equation}
  p[\chi, \Phi^\text{eq}] = \exp\left( - I[\chi] - W[0] \right).
\end{equation}
Expanding to linear order in $\Phi_\alpha-\Phi_\alpha^\text{eq}$ yields
\begin{equation}
  p[\chi,\Phi] = p[\chi, \Phi^\text{eq}] + \frac{\delta^2 \Gamma[\Phi^\text{eq}]}{\delta\Phi_\alpha\delta\Phi_\beta} \, (\Phi_\alpha - \Phi_\alpha^\text{eq}) \, (\phi_\beta[\chi] - \Phi_\beta^\text{eq}) \, p[\chi, \Phi^\text{eq}].
  \label{eq:ProbabilityDensitMixtureClass}
\end{equation}
The interesting statement is that this is not only the linear order of an expansion, but actually an exact expression! The probability density in \eqref{eq:ProbabilityDensitMixtureClass} is indeed constructed such that eq.\ \eqref{eq:ProbabilityDensitMixtureClass} is fulfilled. First note that the distribution is properly normalized, as a result of
\begin{equation}
  \int D \chi \, p[\chi, \Phi^\text{eq}] = 1,
\end{equation}
and
\begin{equation}
  \int D\chi \, \phi_\beta[\chi] \, p[\chi, \Phi^\text{eq}] = \Phi_\beta^\text{eq}.
\end{equation}
Moreover, by construction
\begin{equation}
  \int D\chi \, (\phi_\alpha[\chi] - \Phi_\alpha^\text{eq}) \, (\phi_\beta[\chi] - \Phi_\beta^\text{eq}) \, p[\chi, \Phi^\text{eq}] = G_{\alpha\beta}[J]{\big |}_{J=0}.
\end{equation}
One also needs that $\Gamma^{(2)}$ and $W^{(2)}$ are inverse, as a consequence of the definition \eqref{eq:DefEffectiveAction},
\begin{equation}
  \frac{\delta^2 \Gamma[\Phi]}{\delta \Phi_\alpha \delta\Phi_\beta} \frac{\delta^2 W[J]}{\delta J^\beta \delta J^\gamma} = \frac{\delta^2 \Gamma[\Phi]}{\delta \Phi_\alpha \delta\Phi_\beta} G_{\beta\gamma}[J] = \delta^\alpha_{~\gamma}.
\label{eq:GammaTwoWTwoRelation}
\end{equation}
Together this implies indeed that the distribution in \eqref{eq:ProbabilityDensitMixtureClass} has the expectation values $\Phi_\alpha$.

The class of probability distributions in eq.\ \eqref{eq:ProbabilityDensitMixtureClass}, written as a linear combination of expectation values, is called mixture family in the context of information geometry \cite{amariInformationGeometryIts2016,ayInformationGeometry2017}.

\section{Affine geometry for expectation values}
\label{sec:AffineGeometryForExpectationValues}

The Fisher metric in terms of expectation value coordinates is easily determined to be
\begin{equation}
  G^{\alpha\beta}[\Phi] = - \int D\chi \, p[\chi, \Phi] \,  \frac{\delta^2}{\delta\Phi_\alpha\delta\Phi_\beta} \ln p[\chi, \Phi] = \frac{\delta^2 \Gamma[\Phi]}{\delta \Phi_\alpha\delta\Phi_\beta}.
\end{equation}
As a matrix, this is actually the inverse of $G_{\alpha\beta}[J]$, see eq.\ \eqref{eq:GammaTwoWTwoRelation}. The Fisher-Rao distance between close by distributions can thus be written in three equivalent ways,
\begin{equation}
  d s^2 = G_{\alpha\beta}[J] \delta J^\alpha \delta J^\beta = G^{\alpha\beta}[\Phi] \delta \Phi_\alpha \delta\Phi_\beta = \delta J^\alpha \delta \Phi_\beta.
\end{equation}

As discussed before, the expectation values $\Phi_\alpha$ are related to the sources $J^\alpha$ by a change of variables that is in general nonlinear. In this sense it is surprising that there is an affine structure here, as well. Indeed, transformations of the form
\begin{equation}
  \Phi_\alpha \to \Phi_\alpha^\prime = N_\alpha^{~\beta} \Phi_\beta + d_\alpha,
\end{equation}
with invertible $N_\alpha^{~\beta}$, map elements of the mixture family to elements of the mixture family. One may consider so-called $m$-geodescis connecting two probability distributions with expectation values $\Phi_\alpha^\prime$ and $\Phi_\alpha^{\prime\prime}$ of the form
\begin{equation}
  \Phi_\alpha(t) = (1-t) \Phi_\alpha^\prime + t \Phi_\alpha^{\prime\prime},
  \label{eq:DefMGeodesicEndpoints}
\end{equation}
and the corresponding probability distributions for $0\leq t \leq 1$ are simply linear superpositions of the distributions at the two endpoints of the geodesic.

Note that the $m$-geodesic in \eqref{eq:DefMGeodesicEndpoints} is different from the $e$-geodesic described in \eqref{eq:DefEGeodesicEndpoints}, even when the endpoints $J^{\prime\alpha}$ and $\Phi^\prime_\alpha$ as well as  $J^{\prime\prime\alpha}$ and $\Phi^{\prime\prime}_\alpha$ correspond to the same probability distributions (see also below).

The $m$-geodesics can also be characterized in terms of differential equations,
\begin{equation}
  \frac{d^2}{dt^2} \Phi_\alpha(t) + (\Gamma_\text{M})^{\beta~\gamma}_{\phantom{\beta}\alpha}[\Phi] \, \frac{d}{dt} \Phi_\beta(t) \, \frac{d}{dt} \Phi_\gamma(t) = 0,
\end{equation}
In terms of expectation values as coordinates the $m$-connection symbols vanish, 
\begin{equation}
  (\Gamma_\text{M})^{\beta~\gamma}_{\phantom{\beta}\alpha}[\Phi] = 0. 
\end{equation}

At this point it is instructive to work out the connection symbols of the $m$-connection in terms of source coordinates,
\begin{equation}
  \begin{split}
    (\Gamma_\text{M})_{\alpha~\gamma}^{\phantom{\alpha}\beta}[J] = \frac{\delta \Phi_\alpha}{\delta J^\rho} \frac{\delta \Phi_\gamma}{\delta J^\nu} \frac{\delta J^\beta}{\delta\Phi_\mu} (\Gamma_\text{M})^{\rho\phantom{\mu}\nu}_{\phantom{\rho}\mu}[\Phi] + \frac{\delta J^\beta}{\delta \Phi_\mu} \frac{\delta^2 \Phi_\mu}{\delta J^\alpha \delta J^\gamma}. 
  \end{split}
\end{equation}
This uses the general transformation law for a connection under changes of coordinates. As we just argued, in expectation value coordinates the $m$-connection symbol vanishes, and we thus find in source coordinates
\begin{equation}
  \begin{split}
    (\Gamma_\text{M})_{\alpha~\gamma}^{\phantom{\alpha}\beta}[J] = \frac{\delta J^\beta}{\delta \Phi_\mu} \frac{\delta^2 \Phi_\mu}{\delta J^\alpha \delta J^\gamma} = G^{\beta\mu} \frac{\delta^3 W[J]}{\delta J^\alpha \delta J^\gamma \delta J^\mu}.
  \end{split}
  \label{eq:MConnectionSourceCoordinates}
\end{equation}
Up to a Fisher metric, the $m$-connection symbol in source coordinates is the connected three-point function! Accordingly, $m$-geodesics are in source coordinates not straight lines when this three-point function is non-vanishing.

Similarly, one can find the $e$-connection symbol in expectation value coordinates,
\begin{equation}
  \begin{split}
  (\Gamma_\text{E})^{\alpha\phantom{\beta}\gamma}_{\phantom{\alpha}\beta}[\Phi] = & \frac{\delta J^\alpha}{\delta \Phi_\rho} \frac{\delta J^\gamma}{\delta \Phi_\nu} \frac{\delta \Phi_\beta}{\delta J^\mu} (\Gamma_\text{E})_{\rho\phantom{\mu}\nu}^{\phantom{\rho}\mu}[J] + \frac{\delta \Phi_\beta}{\delta J^\mu} \frac{\delta^2 J^\mu}{\delta \Phi_\alpha \delta \Phi_\gamma} \\
  = & G_{\beta\mu} \frac{\delta^3 \Gamma[\Phi]}{\delta\Phi_\alpha\delta\Phi_\beta\delta\Phi_\mu}.
  \end{split}
  \label{eq:EConnectionExpectationValueCoordinates}
\end{equation}
This is essentially the one-particle irreducible three point function. 

\section{Divergence functional}
\label{sec:DivergenceFunctional}
\subsection{Divergence functional in source coordinates}

Another highly interesting quantity used in information geometry is the Kullback-Leibler divergence or relative information entropy between two probability distributions with coordinates $J^\alpha$ and $J^{\prime\alpha}$,
\begin{equation}
  D[J \| J^\prime] = \int D \chi \, p[\chi, J] \ln\left( p[\chi, J] / p[\chi, J^\prime] \right).
\end{equation}
Note the antisymmetric nature of the definition. The Kullback-Leibler divergence is non-negative,
\begin{equation}
  D[J \| J^\prime] \geq 0,
\end{equation}
and it vanishes when the distributions $p[\chi, J]$ and $p[\chi, J^\prime]$ agree, i.\ e.\ for $J^\alpha = J^{\prime\alpha}$. Moreover, for $J^{\prime\alpha} = J^\alpha + \delta J^\alpha$ one has
\begin{equation}
  D[J \| J^\prime] = \frac{1}{2} G_{\alpha\beta}[J] \delta J^\alpha \delta J^\beta + \ldots,
\end{equation}
where the ellipses are for terms of cubic and higher order in $\delta J^\alpha$. In other words, for nearby distributions the Kullback-Leibler divergence equals the Fisher-Rao distance squared, up to a factor. 

The relative entropy finds an interesting application in terms of Sanov's theorem \cite{demboLargeDeviationsTechniques2010,coverElementsInformationTheory2006}. Consider $n$ random realizations of field configurations $\chi_k$ (where $k=1,\ldots,n$) taken from the distribution \eqref{eq:ProbabilityDistributionExponentialFamily}. This gives rise to an ``empirical'' distribution
\begin{equation}
  p_n[\chi] = \frac{1}{n} \sum_{k=1}^n \delta[\chi - \chi_k].
\end{equation}
The probability $w$ for this empirical distribution $p_n[\chi]$ to lie in some region $\Omega$ of the space of all possible distributions $p[\chi]$ is asymptotically (for large $n$) constrained by Sanov's theorem. For simplicity let us assume that $\Omega$ is closed (with respect to weak topology). Then $w$ is asymptotically constrained by the element $q[\chi]$ of $\Omega$ that is closest to $p[\chi]$ in the sense of relative entropy. In other words, one has
\begin{equation}
  \lim_{n\to \infty} \left( \frac{1}{n} \ln(w) \right) = - \inf_{q\in \Omega} %D(q[\phi] \| p[\phi])
  \int D \chi \, q[\chi] \ln\left( q[\chi] / p[\chi] \right).
\end{equation}
Assuming now in addition that $\Omega$ consists of a set of distributions $p[\chi,J]$ that is in the exponential class \eqref{eq:defFunctProbDistribution} with $J$ in a set $A$, or expectation value $\Phi$ in the corresponding set $B$, we can write $w$ in terms of the divergence functional,
\begin{equation}
  \lim_{n\to \infty} \left( \frac{1}{n} \ln(w) \right) = - \inf_{J\in  A} D[J \| J^\prime]  = - \inf_{\Phi\in  B} D[\Phi \| \Phi^\prime].
\end{equation}
Here we assumed that the true distribution is $p[\phi, J^\prime] = p[\phi, \Phi^\prime]$. By these arguments it becomes clear that the divergence functional plays a natural role in quantifying the probability for large deviations \cite{demboLargeDeviationsTechniques2010}.

For the exponential family one finds easily the relative entropy
\begin{equation}
  \begin{split}
    D[J \| J^\prime] = & (J^\alpha - J^{\prime\alpha}) \Phi_\alpha - W[J] + W[J^\prime] \\
    = & (J^\alpha - J^{\prime\alpha}) \frac{\delta W[J]}{\delta J^\alpha}  - W[J] + W[J^\prime].
  \end{split}
  \label{eq:DivergenceSourceOnly}
\end{equation}
The expectation value $\Phi_\alpha$ is here with respect to the distribution $p[\chi, J]$, as made explicit in the second line. 

The expression in eq.\ \eqref{eq:DivergenceSourceOnly} is up to an interchange of arguments (a duality transform), the Bregman divergence associated with the convex function $W[J]$,
\begin{equation}
  D[J \| J^\prime] = D_W[J^\prime \| J].
\end{equation}
In this context it is also clear that the Schwinger functional $W[J]$ is uniquely fixed by $D[J \| J^\prime]$, up to an additive constant and a linear term in $J$. In other words, to fully reconstruct $W[J]$ from $D[J \| J^\prime]$ from one needs additional information such as $W[0]$ and
\begin{equation}
  \frac{\delta W[J]}{\delta J^\alpha}{\Big |}_{J=0} = \Phi_\alpha^\text{eq}.
  \label{eq:PhieqfromW}
\end{equation}
The largest part of the functional information of $W[J]$ is contained in $D[J \| J^\prime]$. Concretely, one finds for $J^\alpha=0$
\begin{equation}
  D[0 \| J^\prime] + J^{\prime\alpha} \Phi_\alpha^\text{eq} + W[0] = W[J^\prime].
\end{equation}

First derivatives of the divergence functional with respect to the source coordinates are
\begin{equation}
  \begin{split}
    \frac{\delta}{\delta J^\alpha} D[J\| J^\prime] = & (J^\lambda - J^{\prime\lambda} ) \frac{\delta^2}{\delta J^\alpha\delta J^\lambda} W[J] = (J^\lambda - J^{\prime\lambda} ) G_{\alpha\lambda}[J], \\
    \frac{\delta}{\delta J^{\prime\alpha}} D[J\| J^\prime] = & - \frac{\delta}{\delta J^\alpha} W[J] + \frac{\delta}{\delta J^{\prime\alpha}}W[J^\prime] = - \Phi_\alpha[J] + \Phi^\prime_\alpha[J^\prime],
  \end{split}
  \label{eq:firstFunctionalDerivativesSJJ}
\end{equation}
and they vanish for $J=J^\prime$, as expected. Second derivatives are
\begin{equation}
  \begin{split}
    \frac{\delta^2}{\delta J^\alpha\delta J^\beta} D[J\| J^\prime] = & (J^\lambda - J^{\prime\lambda}) \frac{\delta^3}{\delta J^\alpha\delta J^\beta \delta J^\lambda} W[J] + G_{\alpha\beta}[J],\\
    \frac{\delta^2}{\delta J^\alpha \delta J^{\prime\beta}} D[J \| J^\prime] = & - G_{\alpha\beta}[J], \\
    \frac{\delta^2}{\delta J^{\prime\alpha}\delta J^{\prime\beta}} D[J \| J^\prime] = &  G_{\alpha\beta}[J^\prime].
  \end{split}
  \label{eq:secondFunctionalDerivativesSJJ}
\end{equation}
Let us also give the third derivatives,
\begin{equation}
  \begin{split}
    \frac{\delta^3}{\delta J^\alpha \delta J^\beta \delta J^\gamma} D[J \| J^\prime] = & (J^\lambda - J^{\prime\lambda}) \frac{\delta^4}{\delta J^\alpha \delta J^\beta \delta J^\gamma \delta J^\lambda} W[J] + 2 \frac{\delta^3}{\delta J^\alpha \delta J^\beta \delta J^\gamma} W[J], \\
    \frac{\delta^3}{\delta J^\alpha\delta J^\beta\delta J^{\prime\gamma}} D[J \| J^\prime] = & - \frac{\delta^3}{\delta J^\alpha \delta J^\beta \delta J^\gamma} W[J], \\
    \frac{\delta^3}{\delta J^\alpha \delta^{\prime\beta}\delta J^{\prime\gamma}} D[J \| J^\prime] = & 0, \\
    \frac{\delta^3}{\delta J^{\prime\alpha}\delta J^{\prime\beta} \delta J^{\prime\gamma}} D[J \| J^\prime] = & \frac{\delta^3}{\delta J^{\prime\alpha}\delta J^{\prime\beta} \delta J^{\prime\gamma}} W[J^\prime].
  \end{split}
\end{equation}
It is particularly interesting to observe that the derivatives with respect to the second argument $J^\prime$ yield directly the connected correlation functions generated by the Schwinger functional $W[J^\prime]$. Similarly, one derivative with respect to $J^\prime$ and several derivatives with respect to $J$ also lead to connected correlation functions. Schematically one has for $n\geq 2$
\begin{equation}
  D^{(0,n)}[J \| J^\prime] = W^{(n)}[J^\prime],
\end{equation}
and
\begin{equation}
  D^{(n-1,1)}[J \| J^\prime] = - W^{(n)}[J].
\end{equation}
This shows again how most of the information of $W[J]$ is contained in $D[J \| J^\prime]$. 

\subsection{Divergence functional in expectation value coordinates}
The divergence functional can also be expressed in terms of expectation value coordinates, and it is convenient to work then with the quantum effective action defined in \eqref{eq:DefEffectiveAction}. One obtains from \eqref{eq:DivergenceSourceOnly} immediately
\begin{equation}
  D[\Phi \| \Phi^\prime] = \Gamma[\Phi] - \Gamma[\Phi^\prime] - \frac{\delta \Gamma[\Phi^\prime]}{\delta\Phi^\prime_\lambda} (\Phi_\lambda - \Phi^\prime_\lambda).
  \label{eq:SPhiPhipDef}
\end{equation}
This can be seen as the Bregman divergence accociated to $\Gamma[\Phi]$,
\begin{equation}
  D[\Phi \| \Phi^\prime] = D_\Gamma[\Phi \| \Phi^\prime].
\end{equation}
On the other side one can reconstruct $\Gamma[\Phi]$ up to an additive constant from $D[\Phi \| \Phi^\prime]$ if additionally also $\Phi^\text{eq}$ satisfying \eqref{eq:PhieqfromW} or \eqref{eq:FieldEquationEffectiveAction} for $J^\alpha=0$ is known. Specifically, the difference of the quantum effective action at $\Phi$ to the one at $\Phi^\text{eq}$ is the divergence functional,
\begin{equation}
  \Gamma[\Phi] - \Gamma[\Phi^\text{eq}]= D[\Phi \| \Phi^{\text{eq}}]. 
\end{equation}

On similar ground, let us note that from the functional $D[J \| J^\prime]$ alone one cannot immediately infer the functional $D[\Phi \| \Phi^\prime]$ (and vice versa) because the information about the relation between the source $J^\alpha$ and the expectation value $\Phi_\alpha$ is not contained in $D[J \| J^\prime]$ (or in $D[\Phi \| \Phi^\prime]$). It is sufficient, however, to know in addition the expectation value $\Phi_\alpha^\text{eq}$. From \eqref{eq:firstFunctionalDerivativesSJJ} follows a relation for the expectation value as functional of the source,
\begin{equation}
  \Phi_\alpha[J] = \Phi^\text{eq}_\alpha - \frac{\delta}{\delta J^{\prime\alpha}} D[J\| J^\prime] {\big |}_{J^\prime=0},
\end{equation}
which allows to implement the change of variables. Similarly, this can be done in the other direction.

Taking functional derivatives of \eqref{eq:SPhiPhipDef} with respect to the two arguments yields
\begin{equation}
  \begin{split}
    \frac{\delta}{\delta \Phi_\alpha} D[\Phi \| \Phi^\prime ] = & \frac{\delta \Gamma[\Phi]}{\delta \Phi_\alpha} - \frac{\delta \Gamma[\Phi^\prime]}{\delta\Phi^\prime_\alpha} = J^\alpha - J^{\prime\alpha}, \\
    \frac{\delta}{\delta \Phi_\alpha^\prime} D[\Phi \| \Phi^\prime ] = & - \frac{\delta^2 \Gamma[\Phi^\prime]}{\delta \Phi_\alpha^\prime \delta \Phi^\prime_\lambda} (\Phi_\lambda - \Phi_\lambda^\prime) = - G^{\alpha\lambda}[\Phi^\prime] (\Phi_\lambda - \Phi^\prime_\lambda),
  \end{split}
\end{equation}
and they vanish for $\Phi=\Phi^\prime$ as expected. Evaluating the first line at $\Phi^\prime=\Phi^\text{eq}$ leads to a relation for the source as a functional of the expectation value, 
\begin{equation}
  J^\alpha[\Phi] = \frac{\delta}{\delta \Phi_\alpha} D[\Phi \| \Phi^\text{eq} ].
\end{equation}

Second derivatives are
\begin{equation}
  \begin{split}
    \frac{\delta^2}{\delta\Phi_\alpha\delta\Phi_\beta} D[\Phi \| \Phi^\prime] = & G^{\alpha\beta}[\Phi], \\
    \frac{\delta^2}{\delta\Phi_\alpha\delta\Phi_\beta^\prime} D[\Phi \| \Phi^\prime] = & - G^{\alpha\beta}[\Phi^\prime], \\
    \frac{\delta^2}{\delta\Phi_\alpha^\prime\delta\Phi_\beta^\prime} D[\Phi \| \Phi^\prime] = & - \frac{\delta^3}{\delta\Phi^\prime_\alpha\delta^\prime\Phi_\beta \delta\Phi^\prime_\lambda} \Gamma[\Phi^\prime] (\Phi_\lambda - \Phi^\prime_\lambda) + G^{\alpha\beta}[\Phi^\prime],
  \end{split}
\end{equation}
and third derivatives follow as
\begin{equation}
  \begin{split}
    \frac{\delta^3}{\delta\Phi_\alpha\delta\Phi_\beta\delta\Phi_\gamma} D[\Phi \| \Phi^\prime] = & \frac{\delta^3}{\delta\Phi_\alpha\delta\Phi_\beta\delta\Phi_\gamma} \Gamma[\Phi], \\
    \frac{\delta^3}{\delta\Phi_\alpha\delta\Phi_\beta\delta\Phi_\gamma^\prime} D[\Phi \| \Phi^\prime] = & 0, \\
    \frac{\delta^3}{\delta\Phi_\alpha\delta\Phi_\beta^\prime\delta\Phi_\gamma^\prime} D[\Phi \| \Phi^\prime] = & - \frac{\delta^3}{\delta\Phi_\alpha^\prime\delta\Phi_\beta^\prime\delta\Phi_\gamma^\prime} \Gamma[\Phi^\prime], \\
    \frac{\delta^3}{\delta\Phi^\prime_\alpha\delta\Phi^\prime_\beta\delta\Phi^\prime_\gamma} D[\Phi \| \Phi^\prime] = & - \frac{\delta^4}{\delta\Phi^\prime_\alpha\delta\Phi^\prime_\beta\delta\Phi^\prime_\gamma\delta\Phi^\prime_\lambda} \Gamma[\Phi^\prime] (\Phi_\lambda - \Phi_\lambda^\prime) + 2 \frac{\delta^3}{\delta\Phi^\prime_\alpha\delta\Phi^\prime_\beta\delta\Phi^\prime_\gamma}\Gamma[\Phi^\prime].
  \end{split}
  \label{eq:thirdFunctionalDerivativesSPhiPhi}
\end{equation}
Here it is particularly interesting that the derivatives with respect to the first argument $\Phi_\alpha$ yield the one-particle irreducible correlation functions generated by $\Gamma[\Phi]$! Something similar happen for one derivative with respect to $\Phi$ and several with respect to $\Phi^\prime$. Schematically, for $n\geq 2$,
\begin{equation}
  D^{(n,0)}[\Phi \| \Phi^\prime] = \Gamma^{(n)}[\Phi],
\end{equation}
and
\begin{equation}
  D^{(1,n-1)}[\Phi \| \Phi^\prime] = - \Gamma^{(n)}[\Phi^\prime].
\end{equation}
To summarize, the divergence functional $D[\Phi\| \phi^\prime]$, when supplemented with one expectation value configuration $\phi^\text{eq}_\alpha$ corresponding to vanishing source $J^\alpha$, contains the same information as the effective action $\Gamma[\Phi]$.

\subsection{Divergence functional in mixed representation}

There is also a very elegant mixed representation, where the expectation value coordinate is used for the first argument, and the source coordinate for the second argument of the divergence functional,
\begin{equation}
  D[\Phi \| J^\prime] = \Gamma[\Phi] + W[J^\prime] - J^{\prime\alpha} \Phi_\alpha.
  \label{eq:DivergenceFunctionalMixedRepDef}
\end{equation}
Here it is manifest that functional derivatives with respect to the first argument generate one-particle irreducible correlation functions, and those with respect to the second argument generate connected correlation functions. Also, by setting one of the arguments to zero one obtains the quantum effective action or Schwinger functional up to an additive constant, respectively.x

First derivatives are
\begin{equation}
  \begin{split}
    \frac{\delta}{\delta \Phi_\alpha} D[\Phi \| J^\prime] = & \frac{\delta \Gamma[\Phi]}{\delta \Phi_\alpha} - J^{\prime\alpha} = J^\alpha - J^{\prime\alpha}, \\
    \frac{\delta}{\delta J^{\prime\alpha}} D[\Phi \| J^\prime] = & \frac{\delta W[J^\prime]}{\delta J^{\prime\alpha}} - \Phi_\alpha = \Phi^{\prime}_\alpha - \Phi_{\alpha},
  \end{split}
  \label{eq:DivergenceFunctionalMixedRep}
\end{equation}
and they vanish for $J^\alpha=J^{\prime\alpha}$ and $\Phi_\alpha=\Phi^\prime_\alpha$. In particular, when the first line is evaluated at $J^\prime=0$ one obtains
\begin{equation}
  J^\alpha[\Phi] = \frac{\delta}{\delta \Phi_\alpha} D[\Phi \| 0] = \frac{\delta \Gamma[\Phi]}{\delta \Phi_\alpha},
\end{equation}
which is the field equation for the expectation value. Similarly, the second line in \eqref{eq:DivergenceFunctionalMixedRep} gives $\Phi^\prime_\alpha[J^\prime]$ when evaluated at some given configuration $\Phi_\alpha$. In this sense the mixed functional $D[\Phi \| J^\prime]$ contains more information that either $D[J \| J^\prime]$ or $D[\Phi \| \Phi^\prime]$, because also the relation between sources and expectation values is directly contained! Note that $D[\Phi \| J^\prime]$ is not in the form of a Bregman divergence.

Second derivatives are
\begin{equation}
  \begin{split}
    \frac{\delta^2}{\delta\Phi_\alpha\delta\Phi_\beta} D[\Phi \| J^\prime] = & G^{\alpha\beta}[\Phi], \\
    \frac{\delta^2}{\delta\Phi_\alpha\delta J^{\prime\beta}} D[\Phi \| J^\prime] = & - \delta^\alpha_{~\beta}, \\
    \frac{\delta^2}{\delta J^{\prime\alpha}\delta J^{\prime\beta}} D[\Phi \| J^\prime] = & G_{\alpha\beta} [J^\prime].
  \end{split}
\end{equation}
Third derivatives are simply
\begin{equation}
  \begin{split}
    \frac{\delta^3}{\delta\Phi_\alpha\delta\Phi_\beta\delta\Phi_\gamma} D[\Phi \| J^\prime] = & \frac{\delta^3}{\delta\Phi_\alpha\delta\Phi_\beta\delta\Phi_\gamma} \Gamma[\Phi], \\
    \frac{\delta^3}{\delta\Phi_\alpha\delta\Phi_\beta\delta J^{\prime\gamma}} D[\Phi \| J^\prime] = & 0, \\
    \frac{\delta^3}{\delta\Phi_\alpha \delta J^{\prime\beta}\delta J^{\prime\gamma}} D[\Phi \| J^\prime] = & 0, \\
    \frac{\delta^3}{\delta J^{\prime\alpha}\delta J^{\prime\beta}\delta J^{\prime\gamma}} D[\Phi \| J^\prime] = & \frac{\delta^3}{\delta J^{\prime\alpha}\delta J^{\prime\beta}\delta J^{\prime\gamma}} W[J^\prime].
  \end{split}
\end{equation}
More general, one finds here for $n\geq 2$,
\begin{equation}
  \begin{split}
    D^{(n,0)}[\Phi \| J^\prime] = & \Gamma^{(n)}[\Phi], \\
    D^{(0,n)}[\Phi \| J^\prime] = & W^{(n)}[J^\prime]. 
  \end{split}
\end{equation}
Let us summarize that the divergence functional $D[\Phi \| J^\prime]$ in the mixed representation of expectation value for the first argument and source for the second argument contains information equivalent to the Schwinger functional $W[J]$ and effective action $\Gamma[\Phi]$, respectively. In fact, it elegantly combines the two functionals.

\subsection{Divergence functional in opposite mixed representation}
Finally, let us discuss also a mixed representation with the opposite choice of coordinates
\begin{equation}
  D[J \| \Phi^\prime] = J^\lambda \frac{\delta W[J]}{\delta J^\lambda}  - W[J] + \Phi_\lambda^\prime \frac{\delta \Gamma[\Phi^\prime]}{\delta\Phi_\lambda^\prime}  - \Gamma[\Phi^\prime] - \frac{\delta W[J]}{\delta J^\lambda} \frac{\delta \Gamma[\Phi^\prime]}{\delta\Phi^\prime_\lambda}.
  \label{eq:DivergenceFunctionalOppositeMixedRep}
\end{equation}
First derivatives are here
\begin{equation}
  \begin{split}
    \frac{\delta}{\delta J^\alpha} D[J \| \Phi^\prime] = & (J^\lambda - J^{\prime\lambda}) G_{\alpha\lambda}[J], \\
    \frac{\delta}{\delta\Phi^\prime_\alpha} D[J \| \Phi^\prime] = & - G^{\alpha\lambda}[\Phi^\prime] (\Phi_\lambda - \Phi^\prime_\lambda).
  \end{split}
  \label{eq:FirstDerivativeOppositeMixed}
\end{equation}

On first sight $D[J \| \Phi^\prime]$ in this opposite mixed representation seems to contain even less information than $D[J \| J^\prime]$ or $D[\Phi \| \Phi^\prime]$, because even with the additional knowledge of the expectation value $\Phi_\alpha^\text{eq}$ corresponding to vanishing source $J^\alpha=0$, it is not possible to infer from \eqref{eq:FirstDerivativeOppositeMixed} the general relation between expectation values and sources. Only if in addition also the two point function or Fisher metric $G_{\alpha\lambda}[0]$ at $J=0$ or its inverse $G^{\lambda\alpha}[\Phi^\text{eq}]$ is known, can one evaluate the first line of \eqref{eq:FirstDerivativeOppositeMixed} at $J=0$ to yield
\begin{equation}
  J^{\prime\lambda}[\Phi^\prime] = - G^{\lambda\alpha}[\Phi^\text{eq}] \frac{\delta}{\delta J^\alpha} D[J \| \Phi^\prime] {\big |}_{J=0}. 
\end{equation}
Similarly, evaluating the second line at $\Phi^\prime_\alpha = \Phi_\alpha^\text{eq}$ gives
\begin{equation}
  \Phi_\lambda[J] = \Phi_\lambda^\text{eq} - G_{\lambda\alpha}[0] \frac{\delta}{\delta\Phi^\prime_\alpha} D[J \| \Phi^\prime]{\big |}_{\Phi^\prime = \Phi^\text{eq}}.
\end{equation}
However, the necessary information about the Fisher metric is contained in the second functional derivatives of \eqref{eq:DivergenceFunctionalOppositeMixedRep}, which follow in general as
\begin{equation}
  \begin{split}
    \frac{\delta^2}{\delta J^\alpha \delta J^\beta} D[J \| \Phi^\prime] = & (J^\lambda - J^{\prime\lambda}) \frac{\delta^3}{\delta J^\alpha\delta J^\beta \delta J^\lambda} W[J] + G_{\alpha\beta}[J], \\
    \frac{\delta^2}{\delta J^\alpha\delta \Phi^\prime_\beta} D[J \| \Phi^\prime] = & - G_{\alpha\lambda}[J] G^{\lambda\beta}[\Phi^\prime], \\
    \frac{\delta^2}{\delta \Phi^\prime_\alpha \delta \Phi^\prime_\beta} D[J \| \Phi^\prime] = & - \frac{\delta^3}{\delta \Phi^\prime_\alpha \delta \Phi^\prime_\beta \delta\Phi^\prime_\lambda} \Gamma[\Phi^\prime] (\Phi_\lambda-\Phi^\prime_\lambda) + G^{\alpha\beta}[\Phi^\prime].
  \end{split}
\end{equation}
Evaluating the first and third line at $J=0$ and $\Phi^\prime = \Phi^\text{eq}$ yields $G_{\alpha\beta}[0]$ and $G^{\alpha\beta}[\Phi^\text{eq}]$, respectively.

Finally, third derivatives are in this representation given by
\begin{equation}
  \begin{split}
    \frac{\delta^3}{\delta J^\alpha \delta J^\beta \delta J^\gamma} D[J \| \Phi^\prime] = & (J^\lambda - J^{\prime\lambda}) \frac{\delta^4}{\delta J^\alpha \delta J^\beta \delta J^\gamma \delta J^\lambda} W[J] + 2 \frac{\delta^3}{\delta J^\alpha \delta J^\beta \delta J^\gamma}W[J], \\
    \frac{\delta^3}{\delta J^\alpha \delta J^\beta \delta \Phi^\prime_\gamma} D[J \| \Phi^\prime] = & - \frac{\delta^3}{\delta J^\alpha \delta J^\beta \delta J^\lambda} W[J] \frac{\delta^2}{\delta \Phi^\prime_\gamma \delta\Phi^\prime_\lambda} \Gamma[\Phi^\prime], \\
    \frac{\delta^3}{\delta J^\alpha\delta\Phi^\prime_\beta \delta\Phi^\prime_\gamma} D[J \| \Phi^\prime] = & - \frac{\delta^2}{\delta J^\alpha \delta J^\lambda} W[J] \frac{\delta^3}{\delta\Phi^\prime_\beta\delta\Phi^\prime_\gamma\delta\Phi^\prime_\gamma}\Gamma[\Phi^\prime], \\
    \frac{\delta^3}{\delta\Phi^\prime_\alpha\delta\Phi^\prime_\beta \delta\Phi^\prime_\gamma} D[J \| \Phi^\prime] = & - \frac{\delta^4}{\delta\Phi^\prime_\alpha\delta\Phi^\prime_\beta\delta\Phi^\prime_\gamma\delta\Phi^\prime_\lambda} \Gamma[\Phi^\prime] (\Phi_\lambda-\Phi^\prime_\lambda) + 2 \frac{\delta^3}{\delta\Phi^\prime\alpha\delta\Phi^\prime\beta\delta\Phi^\prime_\gamma}\Gamma[\Phi^\prime].
  \end{split}
\end{equation}
This representation is less elegant and obviously leads to more involved expressions for derivatives. 

The question arises to which extent one can recover the Schwinger functional $W[J]$ or quantum effective action $\Gamma[\Phi]$ from the opposite mixed representation functional \eqref{eq:DivergenceFunctionalOppositeMixedRep}. As we have argued, one can, with the additional information of $\Phi^\text{eq}_\alpha$ construct the map between expectation values and sources, and the information in $D[J \| \Phi^\prime]$ is then equivalent to the divergence functional in other representations.

\section{Functional integral representation for divergence functional}
\label{sec:FunctionalIntegralRepresentationForDivergenceFunctional}

For some purposes it is useful to have a direct functional integral representation of functionals. From eq.\ \eqref{eq:DivergenceSourceOnly} one finds
\begin{equation}
  e^{-D[J \| J^\prime]} = \frac{e^{W[J]-J^\alpha\Phi_\alpha}}{e^{W[J^\prime]-J^{\prime\alpha}\Phi_\alpha}}.
\end{equation}
Here one can use the functional integral representation for the Schwinger functional \eqref{eq:SchwingerFunctionalChi}, leading to an intuitive expression for the divergence functional
\begin{equation}
  e^{-D[J \| J^\prime]} = \frac{\int D\chi \, \exp\left( -I[\chi] + J^\alpha (\phi_\alpha[\chi] - \Phi_\alpha) \right)}{\int D\tilde \chi \, \exp\left( -I[\tilde \chi] + J^{\prime\alpha} (\phi_\alpha[\tilde \chi]-\Phi_\alpha) \right)}.
  \label{eq:DivergenceFunctionalIntegralSourcesOnly}
\end{equation}
Note that the expectation value $\Phi_\alpha$ appearing in the numerator, as well as in the denominator is with respect to the distribution at source $J^\alpha$.

At this point it is interesting to undo the coordinate change in \eqref{eq:chiphiCoordinateChange} which yields
\begin{equation}
  e^{-D[J \| J^\prime]} = \frac{\int D\phi \, \exp\left( -S[\Phi] + J^\alpha (\phi_\alpha - \Phi_\alpha)  \right)}{\int D\tilde \phi \, \exp( -S[\tilde \phi] + J^{\prime\alpha} (\tilde \phi_\alpha -\Phi_\alpha) )}.
\end{equation}
Assume now that the two functional integrals can be (approximately) evaluated in steepest descend method (one-loop approximation). This yields
\begin{equation}
  \begin{split}
  D[J \| J^\prime] \approx & \; S[\varphi[J]] - S[\varphi[J^\prime]] - J^\alpha (\varphi_\alpha[J]-\Phi_\alpha) + J^{\prime\alpha} (\varphi_\alpha[J^\prime] - \Phi_\alpha) \\
    & + \frac{1}{2} \text{Tr} \left\{ \ln S^{(2)}[\varphi[J]] - \ln S^{(2)}[\varphi[J^\prime]]\right\},
  \end{split}
  \label{eq:SteepestDescend01}
\end{equation}
where $\phi_\alpha= \varphi_\alpha[J]$ is a solution to the classical field equation
\begin{equation}
  \frac{\delta}{\delta \phi_\alpha} S[\phi] = J^\alpha.
\end{equation}

The functional integral representation \eqref{eq:DivergenceFunctionalIntegralSourcesOnly} can be adapted to other coordinate systems by deriving from \eqref{eq:thirdFunctionalDerivativesSPhiPhi} the relations 
\begin{equation}
  \begin{split}
    J^\alpha = & \frac{\delta}{\delta\Phi_\alpha} D[\Phi \| \Phi^\text{eq}], \\
    J^{\prime\alpha} = & - \frac{\delta}{\delta\Phi_\alpha} D[\Phi \| \Phi^\prime]{\big |}_{\Phi = \Phi^\text{eq}},
  \end{split}
  \label{eq:DeterminationJFromSphiphi}
\end{equation}
where $\Phi^\text{eq}$ is the expectation value configuration corresponding to vanishing source $J=0$. Using this, one finds in terms of expectation value coordinates
\begin{equation}
  e^{-D[\Phi \| \Phi^\prime]} = \frac{\int D\chi \, \exp\left( -I[\chi] + \frac{\delta}{\delta \Phi_\alpha} D[\Phi \| \Phi^\text{eq}] (\phi_\alpha[\chi] - \Phi_\alpha) \right)}{\int D\tilde \chi \, \exp\left( -I[\tilde \chi] - \frac{\delta}{\delta\Phi_\alpha} D[\Phi \| \Phi^\prime]{\big |}_{\Phi=\Phi^\text{eq}} (\phi_\alpha[\tilde \chi]-\Phi_\alpha) \right)}.
  \label{eq:DivergenceFunctionalIntegralExpectationValuesOnly}
\end{equation}
Note that this is an implicit relation, because the divergence functional $D[\Phi \| \Phi^\prime]$ appears on the right-hand side.

For the steepest descend approximation in \eqref{eq:SteepestDescend01} one finds to leading order
\begin{equation}
\begin{split}
  \frac{\delta}{\delta \Phi_\alpha}D[J \| J^\prime] = & \left(\frac{\delta}{\delta\phi_\beta}S[\varphi[J]] - J^\beta \right) \frac{\delta \varphi_\beta[J]}{\delta \Phi_\alpha} - \frac{\delta J^\beta}{\delta \Phi_\alpha} (\varphi_\beta[J] - \Phi_\beta ) + J^\alpha - J^{\prime\alpha} \\
  = & - \frac{\delta J^\beta}{\delta \Phi_\alpha} (\varphi_\beta[J] - \Phi_\beta ) + J^\alpha - J^{\prime\alpha}.
\end{split}
\end{equation}
Eqs.\ \eqref{eq:DeterminationJFromSphiphi} are therefore solved when $\varphi_\beta[J] = \Phi_\beta$. This leads to the leading order steepest descend approximation in expectation value coordinates
\begin{equation}
  \begin{split}
  D[\Phi \| \Phi^\prime] \approx & \; S[\Phi] - S[\Phi^\prime] - \frac{\delta}{\delta \Phi^\prime_\alpha} S[\Phi^\prime] (\Phi_\alpha - \Phi^\prime_\alpha). %\\
  %  & + \frac{1}{2} \text{Tr} \left\{ \ln S^{(2)}[\Phi] - \ln S^{(2)}[\Phi^\prime]\right\}.
  \end{split}
  \label{eq:SteepestDescend02}
\end{equation}
This is consistent with \eqref{eq:SPhiPhipDef} and the standard steepest descend or one-loop approximation for the quantum effective action,
\begin{equation}
  \Gamma[\Phi] \approx S[\Phi] + \frac{1}{2} \text{Tr} \left\{ \ln S^{(2)}[\Phi] \right\}.
\end{equation}

Let us give for completeness also a functional integral relation in the mixed representation \eqref{eq:DivergenceFunctionalMixedRepDef},
\begin{equation}
  e^{-D[\Phi \| J^\prime]} = \frac{\int D\chi \, \exp\left( -I[\chi] + \frac{\delta}{\delta \Phi_\alpha} D[\Phi \| 0] (\phi_\alpha[\chi] - \Phi_\alpha) \right)}{\int D\tilde \chi \, \exp\left( -I[\tilde \chi] + J^{\prime\alpha} (\phi_\alpha[\tilde \chi]-\Phi_\alpha) \right)},
  \label{eq:DivergenceFunctionalIntegralMixed}
\end{equation}
where we used
\begin{equation}
  J^\alpha = \frac{\delta}{\delta \Phi_\alpha} D[\Phi \| J^\prime] {\big |}_{J^\prime=0} = \frac{\delta}{\delta \Phi_\alpha} D[\Phi \| 0] .
\end{equation}
As seen before the mixed representation is particularly elegant in the sense that it does not need additional information about $\Phi^\text{eq}$. 

\section{Connections}
\label{sec:Connections}
Let us now discuss the different connections in more detail. We start by working in source coordinates. Associated to the Fisher metric in eq.\ \eqref{eq:FisherMetricSourceCoordinates} is a unique connection that is both torsion-free and metric-compatible (i.\ e.\ free of non-metricity). This is the Levi-Civita connection
\begin{equation}
  (\Gamma_\text{LC})_{\alpha\phantom{\beta}\gamma}^{\phantom{\alpha}\beta}[J] = \frac{1}{2} G^{\beta\lambda}[J] \left( \frac{\delta}{\delta J^\alpha} G_{\gamma\lambda}[J] + \frac{\delta}{\delta J^\gamma} G_{\alpha\lambda}[J] - \frac{\delta}{\delta J^\lambda} G_{\alpha\gamma}[J]\right).
\end{equation}
For the exponential family one finds easily
\begin{equation}
  (\Gamma_\text{LC})_{\alpha\phantom{\beta}\gamma}^{\phantom{\alpha}\beta}[J] = \frac{1}{2} G^{\beta\lambda}[J] \frac{\delta^3}{\delta J^\alpha\delta J^\beta\delta J^\lambda} W[J],
  \label{eq:LCConnectionSourceCoordinates}
\end{equation}
where $G^{\beta\lambda}[J]$ is the inverse Fisher metric. 

Similarly, the Levi-Civita connection in expectation value coordinates is obtained as
\begin{equation}
  \begin{split}
    (\Gamma_\text{LC})^{\alpha\phantom{\beta}\gamma}_{\phantom{\alpha}\beta}[\Phi] = & \frac{1}{2} G_{\beta\lambda}[\Phi] \left( \frac{\delta}{\delta\Phi_\alpha} G^{\gamma\lambda}[\Phi] + \frac{\delta}{\delta\Phi_\gamma} G^{\alpha\lambda}[\Phi] - \frac{\delta}{\delta\Phi_\lambda} G^{\alpha\gamma}[\Phi] \right) \\
    = & \frac{1}{2} G_{\beta\lambda}[\Phi] \frac{\delta^3}{\delta \Phi_\alpha\delta\Phi_\beta\delta\Phi_\gamma} \Gamma[\Phi],
  \end{split}
\end{equation}
where $G^{\alpha\beta}[\Phi]$ is the Fisher metric in terms of expectation value coordinates and $G_{\alpha\beta}[\Phi]$ its inverse.

Starting from the Levi-Civita connection, one can write any other connection as
\begin{equation}
  \Gamma_{\alpha\phantom{\beta}\gamma}^{\phantom{\alpha}\beta} = (\Gamma_\text{LC})_{\alpha\phantom{\beta}\gamma}^{\phantom{\alpha}\beta} + N_{\alpha\phantom{\beta}\gamma}^{\phantom{\alpha}\beta},
\end{equation}
where $ N_{\alpha\phantom{\beta}\gamma}^{\phantom{\alpha}\beta}$ is known as the distortion tensor. (The difference of two connections transforms as a tensor under coordinate changes or diffeomorphisms.) The torsion tensor can be expressed through the distortion tensor as
\begin{equation}
  T^\alpha_{\phantom{\alpha}\beta\gamma} = N_{\beta\phantom{\alpha}\gamma}^{\phantom{\beta}\alpha} - N_{\gamma\phantom{\alpha}\beta}^{\phantom{\gamma}\alpha},
\end{equation}
and the non-metricity tensor is expressed through the distortion tensor and the metric as
\begin{equation}
  B_{\alpha\beta\gamma} = \frac{1}{2} G_{\beta\lambda} N_{\alpha\phantom{\lambda}\gamma}^{\phantom{\alpha}\lambda} + \frac{1}{2} G_{\gamma\lambda} N_{\alpha\phantom{\lambda}\beta}^{\phantom{\alpha}\lambda}.
\end{equation}

After these general considerations, let us now address the $e$- and $m$-connection. In source coordinates, the $e$-connection symbols vanish, see eq.\ \eqref{eq:EConnectionSourceCoordinates}. Together with eq.\ \eqref{eq:LCConnectionSourceCoordinates} this implies the distortion tensor in source coordinates
\begin{equation}
  (N_\text{E})_{\alpha\phantom{\beta}\gamma}^{\phantom{\alpha}\beta}[J] = -  \frac{1}{2} G^{\beta\lambda}[J] \frac{\delta^3}{\delta J^\alpha\delta J^\beta\delta J^\lambda} W[J].
\end{equation}
Torsion vanishes, and the non-metricity tensor is
\begin{equation}
  (B_\text{E})_{\alpha\beta\gamma}[J] = -  \frac{1}{2} \frac{\delta^3}{\delta J^\alpha\delta J^\beta\delta J^\gamma} W[J].
\end{equation}
Up to a factor, this is simply the connected three-point function!

The $m$-connection symbols in source coordinates are given by eq.\ \eqref{eq:MConnectionSourceCoordinates}. Accordingly, the distortion tensor is here
\begin{equation}
  (N_\text{M})_{\alpha\phantom{\beta}\gamma}^{\phantom{\alpha}\beta}[J] =   \frac{1}{2} G^{\beta\lambda}[J] \frac{\delta^3}{\delta J^\alpha\delta J^\beta\delta J^\lambda} W[J].
\end{equation}
The non-metricity has simply the opposite sign compared to the one of the $e$-connection,
\begin{equation}
  (B_\text{M})_{\alpha\beta\gamma}[J] = - (B_\text{E})_{\alpha\beta\gamma}[J] = \frac{1}{2} \frac{\delta^3}{\delta J^\alpha\delta J^\beta\delta J^\gamma} W[J].
\end{equation}
In the context of information geometry, the fully symmetric tensor $T_{\alpha\beta\gamma} = 2 (B_\text{E})_{\alpha\beta\gamma} = -2(B_\text{M})_{\alpha\beta\gamma}$ is known as the Amari-Chentsov tensor \cite{amariInformationGeometryIts2016,ayInformationGeometry2017}.

By similar arguments, or by a change of coordinates, one also obtains the distortion tensors in expectation value coordinates,
\begin{equation}
  (N_\text{M})^{\alpha\phantom{\beta}\gamma}_{\phantom{\alpha}\beta}[\Phi] = -  (N_\text{E})^{\alpha\phantom{\beta}\gamma}_{\phantom{\alpha}\beta}[\Phi]=   \frac{1}{2} G_{\beta\lambda}[\Phi] \frac{\delta^3}{\delta \Phi_\alpha\delta \Phi_\beta\delta \Phi_\lambda} \Gamma[\Phi].
\end{equation}
The non-metricity tensors are here, up to factors, given by the one-particle irreducible three-point function,
\begin{equation}
  (B_\text{M})^{\alpha\beta\gamma}[\Phi] = -   (B_\text{E})^{\alpha\beta\gamma}[\Phi] = \frac{1}{2} \frac{\delta^3}{\delta \Phi_\alpha\delta \Phi_\beta\delta \Phi_\gamma} \Gamma[\Phi].
\end{equation}

We see that the $e$-connection and $m$-connection are dual in the sense that they have opposite non-metricity and they are both free of torsion. This duality implies that vector fields $V^\mu[J]$ and $W^\nu[J]$ obey
\begin{equation}
  \frac{\delta}{\delta J^\alpha} \left( G_{\mu\nu}[J] V^\mu[J] W^\nu[J]\right) = G_{\mu\nu}[J] \left( \nabla_\alpha^\text{(E)} V^\mu[J] \right) W^\nu[J] + G_{\mu\nu}[J] V^\mu[J] \left( \nabla_\alpha^\text{(M)} W^\nu[J] \right),
  \label{eq:DualtyRelation}
\end{equation}
with the covariant functional derivatives associated with the $e$-connection,
\begin{equation}
  \nabla^\text{(E)}_\alpha V^\mu[J] = \frac{\delta}{\delta J^\alpha} V^\mu[J] + (\Gamma_\text{E})_{\alpha\phantom{\mu}\beta}^{\phantom{\alpha}\mu}[J] \, V^\beta[J] = \nabla^\text{(LC)}_\alpha V^\mu[J] + (N_\text{E})_{\alpha\phantom{\mu}\beta}^{\phantom{\alpha}\mu}[J] \, V^\beta[J], 
\end{equation}
the covariant derivative associated with the $m$-connection,
\begin{equation}
  \nabla^\text{(M)}_\alpha W^\nu[J] = \frac{\delta}{\delta J^\alpha} W^\nu[J] + (\Gamma_\text{M})_{\alpha\phantom{\nu}\beta}^{\phantom{\alpha}\nu}[J] \, W^\beta[J] = \nabla^\text{(LC)}_\alpha W^\nu[J] + (N_\text{M})_{\alpha\phantom{\nu}\beta}^{\phantom{\alpha}\nu}[J] \, W^\beta[J],
\end{equation}
and the covariant derivative of the Levi-Civita connection
\begin{equation}
  \nabla^\text{(LC)}_\alpha V^\mu[J] = \frac{\delta}{\delta J^\alpha} V^\mu[J] + (\Gamma_\text{LC})_{\alpha\phantom{\mu}\beta}^{\phantom{\alpha}\mu}[J] \, V^\beta[J].
\end{equation}
Eq.\ \eqref{eq:DualtyRelation} follows from the well known relation for the Levi-Civita connection
\begin{equation}
  \frac{\delta}{\delta J^\alpha} \left( G_{\mu\nu}[J] V^\mu[J] W^\nu[J]\right) = G_{\mu\nu}[J] \left( \nabla_\alpha^\text{(LC)} V^\mu[J] \right) W^\nu[J] + G_{\mu\nu}[J] V^\mu[J] \left( \nabla_\alpha^\text{(LC)} W^\nu[J] \right),
\end{equation}
and the symmetry properties of the distortion tensors $N_\text{E}$ and $N_\text{M}$.

\section{Metric and connection from divergence}
\label{sec:MetricAndConnectionFormDivergence}
Interestingly it is also possible to obtain the connection symbols directly from the divergence functional \cite{amariInformationGeometryIts2016}. From eqs.\ \eqref{eq:secondFunctionalDerivativesSJJ} one can read off that the metric in source coordinates is given by
\begin{equation}
  G_{\alpha\beta}[J] = - \frac{\delta^2}{\delta J^\alpha \delta J^{\prime\beta}} D[J \| J^\prime]{\big |}_{J=J^\prime}.
\end{equation}
This is an interesting relation because one can do a change of coordinates, e.\ g.\ $J^\alpha \to K^\alpha = K^\alpha[J]$, and both sides transform automatically in the right way,
\begin{equation}
  \begin{split}
  G_{\mu\nu}[K] = & \frac{\delta J^\alpha}{\delta K^\mu} \frac{\delta J^{\prime\beta}}{\delta K^{\prime\nu}} G_{\alpha\beta}[J[K]]  \\
  = & - \frac{\delta J^\alpha}{\delta K^\mu} \frac{\delta J^{\prime\beta}}{\delta K^{\prime\nu}} \frac{\delta}{\delta J^\alpha} \frac{\delta}{\delta J^{\prime\beta}} D[J \| J^\prime]  \\
  = & - \frac{\delta}{\delta K^\mu} \frac{\delta}{\delta K^{\prime\nu}} D[K \| K^\prime]{\big |}_{K=K^\prime}
  \end{split}
\end{equation}
In a related way one can obtain the $m$-connection symbols
\begin{equation}
  \begin{split}
  (\Gamma_\text{M})_{\alpha\beta\gamma}[J] = & G_{\beta\delta}[J] (\Gamma_\text{M})_{\alpha\phantom{\delta}\gamma}^{\phantom{\alpha}\delta}[J] \\
  = & - \frac{\delta^2}{\delta J^\alpha \delta J^\gamma} \frac{\delta}{\delta J^{\prime\beta}} D[J \| J^\prime]{\big |}_{J=J^\prime}.
  \end{split}
\end{equation}
This has indeed the right transformation law for a connection and can immediately be evaluated in any other coordinate system!

Finally, the symbols of the dual $e$-connection vanish in source coordinates, but can be written as
\begin{equation}
  (\Gamma_\text{E})_{\alpha\beta\gamma}[J] = - \frac{\delta}{\delta J^\beta} \frac{\delta^2}{\delta J^{\prime\alpha}\delta J^{\prime\beta}} D[J \| J^\prime]{\big |}_{J=J^\prime}.
\end{equation}
This formula easily generalizes to other coordinates, for example one can obtain \eqref{eq:EConnectionExpectationValueCoordinates} from \eqref{eq:thirdFunctionalDerivativesSPhiPhi}. We observe how elegantly information geometry is encoded in a divergence function.

Let us note here that after a general (non-linear) change of coordinates $J^\alpha \to K^\alpha[J]$, functional derivatives with respect to the sources $J^\alpha$ generalize to covariant derivatives based on the $e$-connetion. One may call them $e$-covariant derivatives. Similarly, functional derivatives with respect to expectation values $\Phi_\alpha$ generalize to covariant $m$-covariant derivatives based on the $m$-connection.

One can develop a calculus for functional derivatives based one these notions of covariant derivates but must be careful with taking over intuition from standard Riemannian geometry in the sense that both covariant derivatives are not metric compatible. One can use the calculus emerging this way to related connected correlation functions to one-particle irreducible correlation functions and vice versa. 

Let us add another remark here. While it is straight forward to express the divergence functional in other coordinates, e.\ g.\
\begin{equation}
  D[K \| K^\prime] = D[J[K] \| J^\prime[K^\prime]],
\end{equation}
it is in general not possible to write the resulting functional $D[K \| K^\prime]$ again in the form of a Bregman divergence with reversed arguments as in \eqref{eq:DivergenceSourceOnly}. This works only when the transformation from $J^\alpha$ to $K^\alpha$ is an affine map.

%\section{Curvature}
Associated with some connection $\Gamma_{\alpha\phantom{\beta}\gamma}^{\phantom{\alpha}\beta}[J]$ is a Riemann curvature tensor
\begin{equation}
  R^{\alpha}_{\phantom{\alpha}\beta\gamma\delta}[J] = \frac{\delta}{\delta J^\gamma}\Gamma_{\delta\phantom{\alpha}\beta}^{\phantom{\delta}\alpha}[J] - \frac{\delta}{\delta J^\delta}\Gamma_{\gamma\phantom{\alpha}\beta}^{\phantom{\gamma}\alpha}[J] + \Gamma_{\gamma\phantom{\alpha}\lambda}^{\phantom{\gamma}\alpha}[J] \Gamma_{\delta\phantom{\lambda}\beta}^{\phantom{\delta}\lambda}[J] - \Gamma_{\delta\phantom{\alpha}\lambda}^{\phantom{\delta}\alpha}[J] \Gamma_{\gamma\phantom{\lambda}\beta}^{\phantom{\gamma}\lambda}[J]. 
\end{equation}
Interestingly, the curvature tensor associated with the $e$- and $m$-connections both vanish! This is immediately clear because $R^{\alpha}_{\phantom{\alpha}\beta\gamma\delta}[J]$ is a tensor, and both the $e$- and the $m$-connection have coordinate systems where their connection symbols vanish. In contrast, the curvature tensor associated with the Levi-Civita connection has no reason to vanish, and it is given by a combination of three points functions and the Fisher metric.

% \section{Quantum field theoretic observables from the divergence functional}

% How to obtain all observables from $D[\Phi \| \Phi^\prime]$ or $D[\Phi \| J^\prime]$ and how to calculate it?

% \section{R\'{e}yni divergence functional}
% It is also interesting to investigate the R\'{e}yni relative entropy, defined in terms of source coordinates as
% \begin{equation}
%   S_\alpha[J \| J^\prime] = \frac{1}{\alpha-1} \ln \left( \int D\chi \, p[\chi, J]^\alpha \, p[\chi, J^\prime]^{1-\alpha} \right).
%   \label{eq:ReyniDivergenceFunctionalSource}
% \end{equation}
% Here $\alpha$ is a parameter, and in the limit $\alpha\to 1$ one obtains the Kullback-Leibler divergence defined in eq.\ \eqref{eq:DivergenceSourceOnly}, $S_1[J \| J^\prime] = D[J \| J^\prime]$. It is straight-forward to see that eq.\ \eqref{eq:ReyniDivergenceFunctionalSource} can be written in terms of the Schwinger functional as
% \begin{equation}
%   S_\alpha[J \| J^\prime] = \frac{1}{\alpha -1} \left( W[\alpha J + (1-\alpha) J^\prime] - \alpha W[J] - (1-\alpha) W[J^\prime] \right).
% \end{equation}
% Interestingly, the argument $\alpha J + (1-\alpha) J^\prime$ is on an $e$-geodesic that connects $J$ with $J^\prime$. 

\section{Conclusions}
\label{sec:Conclusions}

To conclude, we have discussed here the conceptual setup of Euclidean quantum field theory in the functional integral representation from the point of view of information geometry. It is nice to see how naturally the concepts of information geometry apply. When source fields, that are usually introduced to obtain correlation functions, are seen as coordinates, a natural and rich geometric picture arises. Dual to this is a description with field expectation values as coordinates. 

It is clear to any physicist familiar with general relativity how powerful the concepts of differential geometry can be. It is therefore great to have a similar formalism now also available for Euclidean quantum fields. For example one can easily go to general coordinates without losing the significance of convexity of generating functionals. One can work with connections and corresponding covariant derivatives very similar as familiar from spacetime geometry.

A particularly interesting feature of the type of information geometry explored here is that the metric as well as the two dual connections arise from the functional derivatives of a divergence functional. The latter corresponds to the relative information entropy between two probability distributions and it has many highly interesting properties. One of them is that it can serve as a generating functional for correlation functions, very much as the Schwinger functional or the quantum effective action. Another is that it is non-negative and of course it has an information theoretic significance as exemplified by Sanov's theorem.

In the present study we have concentrated on taking the sources or field expectation values as coordinates, but in a very similar way one can also understand coupling constants entering an action as coordinates and extend the information geometry accordingly. This will be done in a forthcoming publication.

Another aspect we did not study here is the renormalization group. In fact, the Schwinger functional and quantum effective action are subject to renormalization. This is discussed in detail in particular in the context of the functional renormalization group \cite{bergesNonPerturbativeRenormalizationFlow2002, pawlowskiAspectsFunctionalRenormalisation2007,giesIntroductionFunctionalRG2012, delamotteIntroductionNonperturbativeRenormalization2012}. In a forthcoming publication we will present a renormalization group flow equation for the divergence functional \cite{floerchingerExactFlowEquation2023}.
 
Finally, information geometry can also be developed for quantum states described by density matrices or reduced density matrices. Central concepts like the relative entropy and Fisher information metric are defined in that context, as well. In the context of relativistic quantum field theory it is particularly interesting that the relative entropy is well-defined also for spatial subregions \cite{arakiRelativeEntropyStates1977}. Relative entropy can be used to formulate thermodynamics \cite{Floerchinger:2020ogh} and relativistic fluid dynamics \cite{Dowling:2020nxc} on the basis of quantum field theory. We can well imagine that a quantum extension of information geometry allows eventually to understand quantum field theory dynamics in much more detail.

\section*{Acknowledgements}
The author would like to thank Holger Gies and Markus Schröfl for useful discussions and Markus Schröfl for carefully reading the manuscript. This work is supported by the Deutsche Forschungsgemeinschaft (DFG, German Research Foundation) under 273811115 – SFB 1225 ISOQUANT.

\providecommand{\href}[2]{#2}\begingroup\raggedright\endgroup

\end{document}